\documentstyle[prl,aps,preprint]{revtex}
\tighten
\input BoxedEPS
\SetRokickiEPSFSpecial  
\HideDisplacementBoxes

\def\mathcal#1{{\cal #1}}
\def\frac#1#2{{#1\over#2}}

\makeatletter
\long\def\@makecaption#1#2{%
\setbox\@testboxa\hbox{\outertabfalse %
\reset@font\small\sf#1\penalty10000\hskip.5em plus.2em\ignorespaces#2%
}%
\setbox\@testboxb\vbox{\hsize 0.8\@capwidth
\ifdim\wd\@testboxa<\hsize %
\hbox to\hsize{\hfil\box\@testboxa\hfil}%
\else %
\small 
\parindent 0pt 
\unhbox\@testboxa\par
\fi
}%
\medskip
\hfil\box\@testboxb\hfil
} %
\def\table{%
\setlength{\columnwidth}{\hsize}
\let\@capwidth  \columnwidth
 \def\@tablenotes{}%
\global\tableontrue
\bgroup\parindent=0pt
\outertabtrue
\setcounter{tablenote}{0}%
\@float{table}%
}%
\def\endtable{%
\global\tableonfalse\global\outertabfalse
{\let\protect\relax\small 
\@tablenotes\par}\xdef\@tablenotes{}%
\end@float\egroup
}%
\@namedef{table*}{%
\let\@capwidth\textwidth 
\def\@tablenotes{}%
\global\tableontrue
\bgroup\parindent=0pt
\outertabtrue
\setcounter{tablenote}{0}%
\@dblfloat{table}%
}
\@namedef{endtable*}{%
\global\tableonfalse\global\outertabfalse
{\let\protect\relax\small\vskip2pt\@tablenotes\par}\xdef\@tablenotes{}%
\end@dblfloat\egroup
}%

\if@floats
\def\figure{\let\@capwidth\columnwidth\@float{figure}}
\let\endfigure\end@float
\@namedef{figure*}{\let\@capwidth\textwidth\@dblfloat{figure}}
\@namedef{endfigure*}{\end@dblfloat}
\else
\def\figure{%
\let\@capwidth\columnwidth
\vskip1pc
\def\@captype{figure}%
\interlinepenalty10000 %
\@ifnextchar[{\@chuckoptarg}{}%
}%
\def\endfigure{\goodbreak\bigskip}%
\@namedef{figure*}{\figure}%
\@namedef{endfigure*}{\endfigure}%
\fi

\makeatother

\begin{document}
\title{Polarized Lambdas in the 
 Current Fragmentation Region\thanks{This work is supported in part by
funds provided by the U.S. Department of Energy (D.O.E.) under cooperative
agreement \#DF-FC02-94ER40818 and \#DE-FG02-92ER40702
and in part by funds provided by the National Science Foundation under grant
\# PHY 92-18167.}}
\author{R. L. Jaffe}
\address{Center for Theoretical Physics, Laboratory for Nuclear Science\\
and Department of Physics\\
Massachusetts Institute of Technology\\
 Cambridge, Massachusetts 02139\\ 
and\\ 
Lyman Physics Laboratory, Harvard University\\
Cambridge, Massachusetts 02138\\
\null}

\date{MIT-CTP-2534\quad
 HUTP-95/A018)\\
 (Submitted to
{\it Phys~Rev~D Rapid Communications}~~May 1996}

\maketitle


\begin{abstract}%
Data on the nucleon spin structure suggests that the $u$ (and $d$)
quark distributions in the $\Lambda$ hyperon may be polarized.  If
this correlation carries over into fragmentation, then the study of
polarized $\Lambda$'s in the current fragmentation region of deep
inelastic lepton scattering will be a sensitive probe of nucleon spin
structure.  $\Lambda$ production by polarized electrons from
unpolarized targets can search for this correlation.  If it is
signficant, $\Lambda$ production by unpolarized electrons from
longitudinally and transversely polarized targets can probe the
$u$-quark helicity and transversity distributions in the nucleon.  We
review what is known about quark polarization in the $\Lambda$,
summarize electroproduction of polarized $\Lambda$s, and estimate the
sensitivity to quark polarizations in the nucleon. We also describe
polarization phenomena associated with vector meson electroproduction
that can be observed in the same experimental configuration.
\end{abstract}

\newpage

In the nonrelativistic quark model all the spin of the $\Lambda$ resides on 
the $s$-quark.  The $u$ and $d$ quarks are supposed to be 
paired to spin and isospin zero.  
The same model predicts that all the spin of the nucleon is carried by its 
quarks.  Data on hyperon $\beta$-decays and 
deep inelastic scattering from polarized nucleons 
shows that the latter is not true.  The latest published
estimates of the spin fraction carried 
by quarks in the nucleon is $\Sigma(10 {\rm GeV}^2) = 0.2\pm 
0.1$ \cite{SMC}.  Applied to the $\Lambda$, the same data indicate that about 
$60\%$ of the $\Lambda$ spin is on $s$ (and $\bar s$) quarks, while $-40\%$ is 
on $u$ (and $\bar u$) and $d$ (and $\bar d$) quarks \cite{BurkJaf}.
These values are as reliable as the values 
of quark spin fractions in the nucleon.

$\Lambda$'s are unique among light hadrons in that their polarization
can be easily reconstructed from the non-leptonic decay
$\Lambda\rightarrow p\pi$.  Other hyperons are too rare to be of much
interest.\footnote{Although $\Sigma^0$'s are common enough to generate a
small depolarizing background 
 via the decay
$\Sigma^0\rightarrow\Lambda\gamma$ \cite{BurkJaf}.  A precision experiment 
should veto $\Lambda$'s secondary to $\Sigma^0$ decay.
}  Other hadrons with spin do
not preserve polarization information in their decay products
because the decays conserve parity.  With the advent of modern deep
inelastic spin physics many authors have examined the potential
role of $\Lambda$'s as a probe of nucleon spin
substructure \cite{Artru,CPR,JafJi,EK,Lu}.  Generally these papers focus only 
on the $s$-quark polarization within the $\Lambda$.  Since polarized 
$s$-quarks are relatively rare in the nucleon and their squared charge is only 
$1/9$, the prospect for using $\Lambda$'s to probe polarized $s$-quarks in the 
nucleon is not too good.  On the other hand, polarized $u$-quarks are abundant 
in the nucleon and their squared charge is $4/9$.  It is easy to see that even 
a small correlation between the spins of the $u$-quarks and the $\Lambda$'s 
into which they fragment would make them dominant over $s$-quarks and 
potentially useful as probes of the polarized $u$-quark distributions in the 
nucleon.\footnote{Note that polarized $s\rightarrow\Lambda$ fragmentation 
functions can play a major role in $e^+e^-\rightarrow\Lambda + X$ on the $Z^0$ 
peak where strange quarks are both copiously produced and strongly 
polarized \cite{Gust,BurkJaf}.}

In this Report I first update what is known about the (integrated) quark 
helicity distributions in the $\Lambda$.  Next, following the work of Artru 
and Mekhfi \cite{Artru}, 
I summarize the opportunities for exploring and exploiting 
$\Lambda$ polarization in electroproduction.  I use a helicity density matrix 
formalism which is particularly simple in the approximation that current 
fragments are produced at small angles in the target rest frame \cite{CGJJ}.
In the last section of this Report
I summarize this formalism and use it explain the spin-dependent effects
accessible through the electroproduction of vector mesons.

My principal conclusions are as follows:
\begin{itemize}
\item Production of $\Lambda$'s in the current fragmentation region by a
{\it longitudinally polarized\/} electron beam scattering off an {\it
unpolarized\/} target has a large analyzing power for the $\Lambda$ helicity
difference fragmentation function $\Delta\hat u_\Lambda(z,Q^2)$.  
$10^4$ events in the current fragmentation region would be sensitive to
$\frac{\Delta\hat u_\Lambda}{\hat u_\Lambda}\approx 0.03$
for the (assumed dominant) $u$-quark.
If the $\Lambda$'s are found to have significant polarization in this 
experiment then the $u\rightarrow\Lambda$ longitudinal spin transfer is likely 
to be the source.
\item If the longitudinally polarized $u\rightarrow\Lambda$
fragmentation function, $\Delta\hat u_\Lambda(z,Q^2)$, is sizeable, then 
production of $\Lambda$'s in the current
fragmentation region by an {\it unpolarized\/} electron beam scattering off a
{\it transversely\/} polarized target may have a
significant sensitivity to the $u$-quark transversity distribution in the 
nucleon.
\end{itemize}
There are many if's  in this project, but the $\Lambda$'s come for free at 
any DIS experiment sensitive to the hadronic final state, such as
the HERMES experiment now underway at DESY \cite{Duren}
and there 
are no candidate experiments now running with greater potential
sensitivity to the nucleon's tranversity.
\section{The Quark Spin Structure of the $\Lambda$}

The quark distribution and fragmentation functions that figure in this 
analysis are defined as follows:

\begin{itemize}
\item $q(x,Q^2)$ is the spin average quark distribution function for flavor $q$,
which contributes to $f_1(x,Q^2) = \frac{1}{2}\sum_q e_q^2 q(x,Q^2)$.
\item $\Delta q(x,Q^2)$ is the helicity difference quark distribution for 
flavor $q$, which contributes to $g_1 (x,Q^2)= 
\frac{1}{2}\sum_q e_q^2\Delta q(x,Q^2)$.
\item $\delta q(x,Q^2)$ is the transversity difference quark distribution,
which contributes to the transversity structure function, $h_1$, best known
for its role in Drell-Yan processes \cite{RS}.  $\delta q$ is identical
to the distribution called $\Delta_T q$ by Artru.
\end{itemize}
We denote antiquark distributions as $\bar q$, $\Delta \bar q$, and 
$\delta \bar q$ respectively.  The fragmentation functions with the same 
spin structure are denoted with a caret: $\hat q$, $\Delta 
\hat q$, and $\delta \hat q$.  Quark distribution or 
fragmentation functions in the nucleon or $\Lambda$ are denoted $q_N$ or 
$q_\Lambda$, respectively, and likewise for fragmentation functions.  
Finally, deep inelastic scattering from polarized nucleons measures the sum of 
quark and antiquark helicity difference distributions which we denote as 
follows:  $\Delta Q\equiv \Delta q + \Delta \bar q$, when necessary.
 
The integrated polarized quark distributions in any octet baryon state are 
determined by the
$\mathcal{F}$ and $\mathcal{D}$ constants from hyperon $\beta$-decay and the 
$g_1$ sum rules measured in deep inelastic scattering from nucleons.  The 
analysis rests on the assumption of $SU(3)$ symmetry for hyperon 
$\beta$-decay, which gives $\mathcal{F}+\mathcal{D}=1.2573\pm 0.0028$ and 
$\mathcal{F}/\mathcal{D}=0.575\pm 0.016$.  
The errors are dominated by the $g_1$ sum rules which we summarize by the quark 
spin fraction, $\Sigma(10 {\rm Gev}^2)= 0.20\pm 0.11$ \cite{SMC}.
The $\Lambda$ spin fractions are given by
\begin{eqnarray}
\Delta U_\Lambda &= &\Delta D_\Lambda = \frac{1}{3}(\Sigma -
\mathcal{D})\nonumber\\
\Delta S_\Lambda &= &\frac{1}{3}(\Sigma + 2\mathcal{D})\label{deltal}
\end{eqnarray}
Naive quark model results are obtained 
by setting $\mathcal{D}=1$, $\mathcal{F}=2/3$, and $\Sigma = 1$.  A
somewhat  more sophisticated 
estimate is obtained by combining the 
measured $\mathcal{F}$ and $\mathcal{D}$ values with the 
assumption of no polarized $s$-quarks in the nucleon, which we call the 
$\Delta S_N = 0$ model \cite{EJ}.  Table~I summarizes the 
information now available on the 
integrated polarized quark distributions in the nucleon and the $\Lambda$.  For 
comparison we tabulate the expected values in the naive quark model and in the 
$\Delta S_N=0$ approximation.
These entries are for pedagogical purposes only --- the important 
entries come from the data.

\def\ph{\phantom{-}}
\newcommand{\STRUT}{\rule{0in}{3ex}}
 \table  
 \small\centering
\hspace*{-1em}\vbox{\def\Strut{\vrule height11pt depth3pt width 0pt}%
\def\STRUT{\vrule height12pt depth6pt width 0pt}%
\def\frac#1#2{\textstyle{#1\over#2}}%
\def\ph{\phantom{-}}
\let\quad\enspace
\offinterlineskip\halign{
\Strut#&
\vrule#\enspace&
\hfil$#$\hfil&
\enspace\vrule#\quad&
\hfil$#$\hfil&
\quad\vrule#\quad&
\hfil$#$\hfil&
\quad\vrule#\quad&\hfil$#$\hfil&
\quad\vrule#\enspace&
\hfil$#$\hfil&
\enspace\vrule#\quad&
\hfil$#$\hfil&
\quad\vrule#\quad&
\hfil$#$\hfil&
\quad\vrule#\quad&\hfil
$#$\hfil\quad\vrule \cr
\noalign{\hrule}
\Strut&&&\omit&\multispan5\hfil Nucleon\hfil&&&\omit&\multispan5\hfil
Lambda\hfil\vrule\cr 
\noalign{\hrule}
&&&& \hbox{Naive QM} && \Delta S_N=0 && \hbox{Data}&&
&&\hbox{Naive QM}&&\Delta S_N=0&&\hbox{Data} \cr
\noalign{\hrule}
\STRUT&&\Delta U_N &&\ph\frac{4}{3} &&\ph0.92\pm 0.04 &&\ph0.79\pm 0.04 &&\Delta
U_\Lambda &&0 &&-0.07\pm 0.01 &&-0.20\pm 0.04 \cr
\noalign{\hrule}
\STRUT&&\Delta  D_N &&-\frac{1}{3} &&-0.34\pm 0.05 &&-0.47\pm 0.04 &&\Delta D_\Lambda
&&0 &&-0.07\pm 0.01 &&-0.20\pm 0.04 \hfill \cr
\noalign{\hrule}
\STRUT&&\Delta S_N &&\ph0 && 0 && -0.12\pm 0.04 &&\Delta S_\Lambda &&1 &&\ph0.72\pm 0.02
&&\ph0.60\pm0.04\hfill
\cr
\noalign{\hrule}}}%
\label{tab:DeltaI}\nopagebreak
\caption{Light quark spin fractions in the nucleon
and $\Lambda$ as  predicted by the naive quark model, by baryon $\beta$-decay plus the
assumption  that there are no polarized $s$-quarks in the nucleon, and as measured.}
\endtable
The message from Table~I is that the $u$- (and $d$-)quarks 
in the $\Lambda$ are polarized.  One might suppose, however, that this has 
little to do with the $\Lambda$ structure and more to do with the sea of 
$Q\bar Q$ pairs present in all baryons.  To analyze this possibility we 
consider two different parameterizations of the sea quarks in the $\Lambda$.  
In both cases we assume that the sea quark polarization distribution in the 
nucleon and $\Lambda$ are identical.  This is a crude approximation necessary 
to obtain a first order estimate.  The failure of the Gottfried relation  
shows that antiquark distributions depend on valence quark content \cite{Gott}.
In Case I we go further and assume that 
the sea is $SU(3)$ flavor symmetric:  $\Delta s_N = \Delta\bar s_N = 
\Delta\bar u_\Lambda = \Delta\bar d_\Lambda = \Delta\bar s_\Lambda$, {\it 
etc.\/}  In particular,
$\Delta\bar u_\Lambda = \Delta s_N = \frac{1}{2} \Delta S_N = -0.06\pm 0.02$ 
(Case I).  Alternatively we suppose that polarized
$s$-quarks are suppressed in the baryon sea.  We 
choose a suppression factor of $1/2$ from the neutrino data on the $s$-quark 
momentum distribution in the nucleon \cite{CCFR}. Then $\Delta\bar u_\Lambda = 
\Delta\bar d_\Lambda = \Delta\bar u_N = \cdots = 2\Delta s_N = \Delta S_N$, 
or $\Delta\bar u_\Lambda = \Delta S_N = -0.12\pm 0.04$ (Case II).

\table
\centering
\hfil\vbox{\def\Strut{\vrule height13pt depth3pt width 0pt}
\offinterlineskip\bigskip
\halign{
\vrule\strut#&\quad\hfil#\hfil\quad&\vrule#&
 \quad\hfil#\hfil\quad&\vrule#&\quad\hfil#\hfil\quad\vrule\cr
\noalign{\hrule}
&Distribution && Case I && Case II\cr
\noalign{\hrule}
\Strut&$\Delta u_\Lambda$ && $-0.14\pm 0.04 $ && $-0.09\pm 0.03 $\cr
\Strut&$\Delta d_\Lambda$ && $-0.14\pm 0.04 $ && $-0.09\pm 0.03 $\cr
\Strut&$\Delta s_\Lambda$ && $\ph0.66\pm 0.04 $ && $\ph0.66\pm 0.04
$\cr
\noalign{\hrule}}}\hfil%
\label{tab:deltaII}\nopagebreak
\caption{\sf Polarized quark and antiquark content of the
$\Lambda$.}
\endtable

Combining the data from Table~I with the two cases we arrive 
at estimates of the polarized quark and (separately) antiquark content of the 
$\Lambda$.  These are summarized in Table~II\null.  In both cases, 
the $u$-quark's spin is (anti)correlated with the $\Lambda$ spin.  
The correlation is relatively small ($-0.14\pm 0.04$ [Case I] or $-0.09\pm 
0.04$ [Case II]), but {\it if it carries over into the polarized fragmentation 
function}, $\Delta \hat u_\Lambda$, then the $u$-quark will dominate 
polarized $\Lambda$ production in deep inelastic scattering.  To illustrate 
this, consider the product $\Pi_q\equiv
\Delta q_N \times e_q^2 \times \Delta \hat q_\Lambda$, which 
roughly measures the importance of the quark of flavor $q$ in polarized 
$\Lambda$ production from nucleons.  Taking $\Delta\hat q_\Lambda \propto 
\Delta q_\Lambda$, 
for Case I, ${\Pi_u}/{\Pi_s}\approx 12$ 
and for Case II, ${\Pi_u}/{\Pi_s}\approx 8$.  The crucial question remains 
whether the correlation between the $u$-quark spin and the $\Lambda$ spin 
persists in fragmentation.  Experiment will tell.

\section{Polarization transfer in deep inelastic scattering}
The analysis of polarization transfer in deep inelastic scattering is made
considerably simpler by the fact that the electron scattering angle is so
small.  At $E_e\approx 30 {\rm GeV}$, $x\approx 0.1$, and $Q^2\approx 3 {\rm 
GeV}^2$, $\theta\approx 0.08$.  Complexities 
in the analysis of fragment polarization
turn out to be proportional to $\sin^2\theta$ and can be ignored at fixed 
target facilities of interest.

In this approximation the production of $\Lambda$'s can be viewed as an
essentially collinear process.  For {\it longitudinally polarized electrons and
an unpolarized target,\/} the crucial question is how effectively is the
electron polarization transferred to the $\Lambda$ fragment.  The answer is
\begin{equation}
\vec\mathcal{P}_\Lambda =\hat e_3 \mathcal{P}_e \frac{y(2-y)}{1+(1-y)^2} 
\frac{\sum_q e_q^2 q_N(x,Q^2) \Delta\hat q_\Lambda(z,Q^2)}{%
 \sum_q e_q^2 q_N(x,Q^2) \hat q_\Lambda(z,Q^2)},
\label{elongpol}
\end{equation}
where, by convention, the electron beam defines the $\hat e_3$ axis, and $y$
is the usual DIS variable, $y\equiv {(E-E')}/{E}$.

For a {\it polarized target and unpolarized beam\/} 
the calculation is somewhat more
complicated but the results simplify considerably in the $\sin\theta\approx
0$ limit.   When the target is polarized along the electron beam, we find
\begin{equation}
\vec\mathcal{P}_\Lambda =\hat e_3 \mathcal{P}_N 
\frac{\sum_q e_q^2 \Delta q_N(x,Q^2) \Delta\hat q_\Lambda(z,Q^2)}{
\sum_q e_q^2 q_N(x,Q^2) \hat q_\Lambda(z,Q^2)},
\label{Nlongpol}
\end{equation}
and when the target is polarized transverse to the electron beam, we find
\begin{equation}
\vec\mathcal{P}_\Lambda =\vec\mathcal{P}_N \frac{2(1-y)}{1+(1-y)^2}
  \frac{\sum_q e_q^2 \delta q_N(x,Q^2) \delta\hat q_\Lambda(z,Q^2)}{%
\sum_q e_q^2 q_N(x,Q^2) \hat q_\Lambda(z,Q^2)},
\label{Ntranspol}
\end{equation}
as first derived by Artru and Mekhfi \cite{Artru}.
Eqs.~(\ref{elongpol}), (\ref{Nlongpol}), and (\ref{Ntranspol}) 
are the fundamental
results here.  They are accurate to leading twist in the small $\theta$
limit.  Also, $R={\sigma_L}/{\sigma_T}$ was set to zero in the derivation.

Returning to eq.(\ref{elongpol}) assuming $u$-quark dominance, we find
\begin{equation}
\vec\mathcal{P}_\Lambda =\hat e_3 \mathcal{P}_e \frac{y(2-y)}{ 1+(1-y)^2} 
\frac{\Delta\hat u(z,Q^2)}{%
 \hat u(z,Q^2)}.
\label{elongu}
\end{equation}
At $E_e\approx 30 {\rm GeV}$, $x=0.1$ and $Q^2=3 {\rm GeV}^2$, $y=0.53$.  
With a beam polarization of 50\% we find
$\mathcal{P}_\Lambda(z,Q^2) = 0.3 \frac{\Delta \hat u_\Lambda(z,Q^2)}{ \hat 
u_\Lambda (z,Q^2}$.\quad $10^4 \ \Lambda$'s in the current
fragmentation regions  would be sensitive to ${\Delta \hat
u_\Lambda}/{ \hat  u_\Lambda}$ as small as $0.03$.

If existing data do show a significant longitudinal $q\rightarrow\Lambda$
spin transfer in the current fragmentation region, then the next step would
be to orient the target spin transverse to an (unpolarized) electron beam and
look for a {\it transverse\/}
$\Lambda$ polarization, which would provide the first measurement of the
$u$-transversity distribution in the nucleon.  In the $u$-quark dominance
approximation eq.(\ref{Ntranspol}) reduces to
\begin{equation}
\vec\mathcal{P}_\Lambda =\vec\mathcal{P}_N \frac{2(1-y)}{1+(1-y)^2}
\frac{ \delta u_N(x,Q^2)}{u_N(x,Q^2)} \frac{\delta\hat u_\Lambda(z,Q^2)}{
\hat u_\Lambda(z,Q^2)}.
\label{Ntransu}
\end{equation}
Neither the transverse $N\rightarrow u$ polarization transfer nor the
transverse $u\rightarrow\Lambda$ polarization transfer are known.  A
dedicated run with transversely polarized target would open the opportunity
for measurement of both of these novel quark spin distributions.  

It should be noted that the polarization described in eq.(\ref{Ntransu}) is
leading twist.  Inclusive transverse polarization phenomena in DIS are
twist-three and consequently difficult to observe.  The dominant, twist-two
polarization distribution, $\delta q$, decouples from {\it inclusive\/} DIS
because it is chiral-odd \cite{JafJi}.  In eq.(\ref{Ntransu}) the chiral-odd
transverse fragmentation function $\delta\hat q$ combines with the
chiral-odd transverse distribution function $\delta q$ to conserve
net quark chirality \cite{Artru}.  
So observation of the final $\Lambda$ transverse spin
acts as a filter that selects a piece of the transverse spin asymmetry
that does not fall like ${1}/{\sqrt{Q^2}}$.  
At twist three there are other processes which may
provide a measure of $\delta q$.  In particular, inclusive pion production
from a transversely polarized nucleon is sensitive to $\delta q$ \cite{JafJi}.
Since pions dominate the final state in the current fragmentation region it is
possible that their abundance may compensate for the suppression of
twist-three by ${1}/{\sqrt{Q^2}}$.  In any case the pion and $\Lambda$
data could be accumulated simultaneously in a dedicated run with a
transversely polarized target.

\section{Density Matrix Formalism and Application to Vector Meson Production}

The formalism used for the calculations summarized in the previous section
is simple and general.  It is based on a helicity formalism that renders the 
parton formulation of spin dependent processes at leading twist essentially 
trivial.  This formalism matches easily onto the helicity amplitude formalism 
in which perturbative QCD calculations are already known to simplify 
considerably \cite{GastWu}.

At leading twist quarks are described by two-component spinor field built out 
of the so-called ``good'' light-cone fields.  We use a basis of helicity 
eigenstates for these fields.\footnote{At next-to-leading twist, a second 
two-component field, built of ``bad'' light-cone components, enters and may be 
treated with similar methods.}  The various quark (and gluon) distribution and 
fragmentation functions can be related to helicity amplitudes, and transcribed 
as helicity density matrices.  If the hard scattering cross sections are 
transcribed to the same basis, then the extraction of spin
dependent observables reduces to multiplication of helicity matrices.  
Parts of this formalism were developed in 
Ref.~\cite{CGJJ} and a review of the underlying light-cone physics can be found 
in Ref.~\cite{erice}.

The basic ingredients are the $N\rightarrow q$ distribution function, the
$q\rightarrow \Lambda$ fragmentation function and the hard partonic cross
section, all as density matrices 
in the helicity basis.  We discuss the $N\rightarrow q$ 
distribution function in detail first.

\begin{figure}[htbp] 
\centerline{\BoxedEPSF{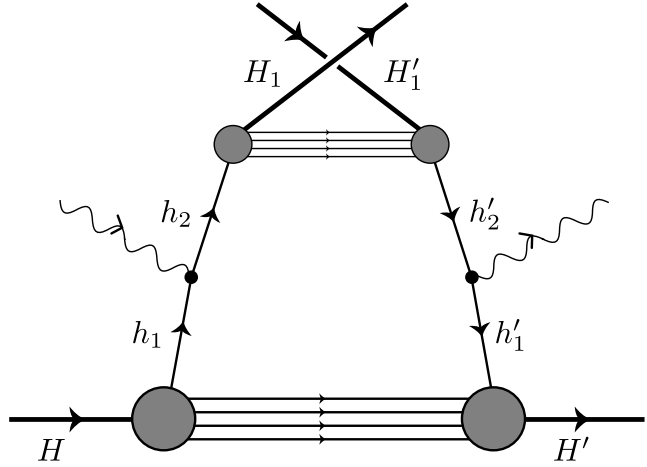 scaled 1200}}
\caption{Hard scattering diagram for $\Lambda$ production in the current 
fragmentation region of lepton scattering 
from a target nucleon.  In perturbative QCD the diagram factors into the 
products of distribution function (lower), hard scattering (middle), and 
fragmentation function (upper part of diagram).  Helicity density matrix 
labels are shown explicitly.}
\label{fig:F}
\end{figure}

The distribution function $\mathcal{F}$
is a function of $x$ and
$Q^2$ carries both quark ($h_1h'_1$) and nucleon ($H'H$) helicity labels.  It
describes the emission of a helicity $h_1$ quark with momentum fraction $x$
by a nucleon of helicity $H$, followed by reabsorption of the quark, with
helicity $h'_1$ forming a nucleon of helicity $H'$.  The process is shown at 
the bottom of 
Fig.~[\ref{fig:F}]  By convention all
quark helicities are lower case, all baryon helicities are upper case, and
helicity matrices carry outgoing particle indices on the left and incoming on
the right.  Thus this distribution function is labelled, $\mathcal{
F}_{H'H,h_1h'_1}$.

Conservation of angular momentum along the $\hat e_3$-axis (defined by the 
problem at hand) restricts the helicities to
\begin{equation}
H + h_1' = H' + h_1.
\label{conservation}
\end{equation}
Parity and time reversal restrict the number of independent components of 
$\mathcal{F}$:
\begin{eqnarray}
\mathcal{F}_{H'H,h_1h'_1} & = & \mathcal{F}_{-H'-H,-h_1-h'_1}\quad {\rm 
	(parity)}\nonumber\\
\mathcal{F}_{H'H,h_1h'_1} & = & \mathcal{F}_{HH',h'_1h_1}\quad {\rm 
	(T-reversal).}
\label{pt}
\end{eqnarray}
Only three independent amplitudes remain,
\begin{eqnarray}
\mathcal{F}_{++,++} & = & \mathcal{F}_{--,--} = q + \Delta q\nonumber\\
\mathcal{F}_{++,--} & = & \mathcal{F}_{--,++} = q - \Delta q\nonumber\\
\mathcal{F}_{+-,-+} & = & \mathcal{F}_{-+,+-} = \delta q
\label{densityF}
\end{eqnarray}
and they can be identified with the conventional quark distributions, 
$q(x,Q^2)$, $\Delta q(x,Q^2)$ and $\delta q(x,Q^2)$ on the basis of the known 
helicity structure of the distributions.  To preserve clarity, the flavor,
$Q^2$, $x$, and target labels on eq.(\ref{densityF}) have been 
suppressed.  The helicity $\pm\frac{1}{2}$ states have been denoted $\pm$ 
respectively.

It is now trivial to encode the information in eq.(\ref{densityF}) 
in a (double) density 
matrix notation.  By inspection \cite{JafJi,CGJJ},
\begin{equation}
\mathcal{F} = q~I\otimes I +  \Delta q~\vec\sigma\cdot \hat e_3 \otimes 
\vec\sigma\cdot\hat e_3+\delta q~
\sum_{j = 1,2}\vec\sigma\cdot \hat e_j \otimes\vec\sigma\cdot\hat e_j .
\label{calf}
\end{equation}
The $\{\sigma^k\}$ are the usual Pauli
matrices.  The first matrix in the direct product $M\otimes N$ is in the
nucleon helicity space, the second is in the quark helicity space.  Thus
$I\otimes I$ denotes $\delta_{H'H}\delta_{h_1h'_1}$.  The dependence of the
distributions ($q$, {\it etc.\/}) on $x$ and $Q^2$ is suppressed.  The
remarkably simple form of $\mathcal{F}$ displays the analogy between
longitudinal and transverse spin effects at leading twist in pQCD.

The fragmentation function is defined analogously.  A quark of helicity
$h_2$ fragments into a $\Lambda$ of helicity $H_1$,
followed by reabsorbing the $\Lambda$ of helicity $H'_1$ to reform a
quark of helicity $h'_2$.  The process is shown at the top of 
Fig.~[\ref{fig:F}].  The
fragmentation function is a distribution differential in the momentum
fraction of the observed hadron: $\left[{d\mathcal{D}/
dz}\right]_{h'_2h_2,H_1H'_1}$, and is given by \cite{CGJJ}:
\begin{equation}
\frac{d\mathcal{D}}{dz} = \frac12\hat q~I\otimes I  + \frac12\Delta \hat q~
\vec\sigma\cdot\hat e'_3 \otimes
\vec\sigma\cdot\hat e'_3 + \frac12\delta \hat q~
\sum_{j=1,2}\vec\sigma\cdot\hat e'_j \otimes \vec\sigma\cdot\hat e'_j.
\label{cald}
\end{equation}

The notation in eqs.~(\ref{calf}) and (\ref{cald}) requires a further word of 
explanation.  In general the axis along which the helicity of the quark 
distribution function is defined ($\hat e_3$) does not coincide with the axis 
along which the helicity of the fragmentation function is defined ($\hat 
e'_3$).  Thus it is important to distinguish the helicity basis vectors in the 
two.  The hard scattering cross section must be a double density 
matrix with one ``leg'' in the unprimed basis and the other in the primed 
basis.  When the distinction between the two can be ignored ({\it e.g.\/} 
$\sin\theta\approx 0$) then $\vec\sigma\cdot \hat e_3 = \vec\sigma\cdot \hat 
e'_3 = \sigma_3$, {\it etc.\/} can be substituted to simplify 
eqs.~(\ref{calf}) and (\ref{cald}).

The cross section for the hard QCD process of interest is obtained by
combining these distribution functions with the appropriate hard parton
scattering density matrix.  In the case of interest here the process is
forward virtual Compton scattering in which an incoming quark of momentum
$p$ and helicity $h_1$ absorbs a virtual photon of momentum $q$ to become an
outgoing quark of momentum $p'=p+q$ and helicity $h_2$.  To allow for the
most general possible spin structure we must consider the process in which
the conjugate amplitude has different helicities, $h'_1$ and
$h'_2$.  This is shown in the middle of Fig.~[\ref{fig:F}].  
The resulting hard density matrix
is denoted $\left[\frac{d\mathcal{M}}{
dx\,dy\,d\phi}\right]_{h_2h'_2,h'_1h_1}$ where
$\phi$ is the azimuthal angle of the scattering plane.  

 Up to inessential kinematic factors $\left[\frac{d\mathcal{M}}{
dx\,dy\,d\phi}\right]_{h_2h'_2,h'_1h_1}$ 
is given by
\begin{eqnarray}
\biggl[\frac{d\mathcal{M}^\pm}{dx\,dy\,d\phi}\biggr]_{h_2h'_2,h'_1h_1}\!\! &= & \bar
u(p,h'_1)\gamma_\mu  u(p',h'_2)\bar u(p',h_2)\gamma_\nu u(p,h_1)\nonumber\\
&{\phantom{=}}&\left( k^\mu k^{\prime\nu} + k^{\prime\mu} k^\nu - g^{\mu\nu}k\cdot k'\mp
i\varepsilon^{\mu\nu\alpha\beta}k_\alpha k'_\beta\right),
\label{sigma}
\end{eqnarray}
where $k$ and $k'$ are the initial and final electron momenta respectively,
$u$ and $\bar u$ are Dirac spinors and the $\pm$ sign refers to the initial
electron helicity.  The resulting density matrix has four terms,
\begin{equation}
\frac{d\mathcal{M}^\pm}{dx\,dy\,d\phi}= A~I\otimes I + B~\vec\sigma\cdot\hat e_3\otimes
\vec\sigma\cdot\hat e'_3 + \sum_{j,\ell = 1,2} C_{j\ell}~\vec\sigma\cdot\hat
e_j\otimes\vec\sigma\cdot\hat e'_\ell\pm D\left(I\otimes
\vec\sigma\cdot\hat e_3' + \vec\sigma\cdot\hat e_3\otimes I\right)
\label{sigmahel}
\end{equation}
The coefficients $A,B,C_{j\ell},D$ are easily calculated once one has an
expression for the Dirac spin density matrix in a helicity basis.  Let
\begin{equation}
U(p)_{hh'}\equiv u(p,h)\bar u(p,h').
\label{U}
\end{equation}
Using the familiar definition of the spin projection operator,
\begin{equation}
\mathcal{P}(p,s)\equiv u(p,s)\bar u(p,s) = \frac{\gamma\cdot p + m}{2m}
\frac{m+\gamma_5\gamma\cdot s}{2}
\label{dirac}
\end{equation}
and taking the $m\rightarrow 0$ limit carefully (maintaining $s^2=-m^2$ and
$s\cdot p = 0$), we find
\begin{equation}
U(p) = \frac{1}{2}\gamma\cdot p\left[ I -\gamma_5(\vec\sigma\cdot\hat e_3
+ \sum_{j=1,2}\vec\gamma_j\cdot\hat e_j \vec\sigma_j\cdot\hat
e_j)\right].
\label{Uh}
\end{equation}
With the aid of eq.(\ref{Uh}) we find  
\begin{eqnarray}
A&=& \frac{e^4}{32\pi^2 Q^2}\frac{1+(1-y)^2}{2y}\nonumber\\
B&=& A\nonumber\\
D&=& \frac{y(2-y)}{1+(1-y)^2}A\nonumber\\
C_{j\ell} &= &\{2(1-y)~ \hat e_j\cdot \hat e'_\ell
+\frac{2}{Q^2}\vec p'\cdot\hat e_j(y \vec{ k'}\cdot\hat e'_\ell + y(1-y)
\vec k\cdot \hat e'_\ell) \nonumber\\
&-&\frac{4y^2}{ Q^2} \vec k'\cdot\hat e_j \vec k\cdot
\hat e'_\ell\} \frac{A}{1+ (1-y)^2}.
\label{sigmaterms}
\end{eqnarray}
Finally, when $\sin\theta$ can be ignored $C_{j\ell}$ simplifies to
\begin{equation}
\lim_{\sin\theta\rightarrow 0} C_{j\ell} = \frac{2(1-y)}{ 1+
(1-y)^2}\delta_{j\ell} A,
\label{Csimple}
\end{equation}
which is the result quoted by Artru and Mekhfi \cite{Artru}.

To obtain the $\Lambda$ production density matrix in the
current fragmentation region for electrons of a given helicity incident on a
target of given spin orientation, it remains only to multiply the
ingredients,
\begin{eqnarray}
\frac{d\mathcal{M}^\pm}{dxdydzd\phi} =
\frac{e^4}{4\pi^2Q^2}&\ &\biggl\{\frac{1+(1-y)^2}{2y}
\sum_q e_q^2 q(x,Q^2)\hat q(z,Q^2)~ I\otimes I\nonumber\\
 &+&\frac{1+(1-y)^2}{2y}\sum_q e_q^2\Delta q(x,Q^2)\Delta\hat q(z,Q^2)~
\sigma^3\otimes\sigma^3\nonumber\\
& +&\frac{2(1-y)}{2y}\sum_q e_q^2 \delta q(x,Q^2)
\delta\hat q(z,Q^2)~\sum_{j=1,2}\sigma^j\otimes\sigma^j\nonumber\\
&\pm& \frac{2-y}{ 2}\sum_q e_q^2 [  q(x,Q^2)
\Delta \hat q(z,Q^2)~I\otimes \sigma^3\nonumber\\
&+ & \Delta q(x,Q^2)\hat q(z,Q^2)~
\sigma^3\otimes I ] \biggr\}
\label{final}
\end{eqnarray}
$\mathcal{M}$ is a matrix in the helicity space of both the nucleon and
the $\Lambda$.  The first (second) matrix lies in the nucleon ($\Lambda$) 
helicity basis.  Note that the $\hat e_j$ and $\hat e'_j$
bases have become identical in the $\sin\theta\rightarrow 0$ limit.
Given $\mathcal{M}$, the
$\Lambda$ polarization is defined by $\frac{{\rm
Tr}\{\mathcal{M}\vec\sigma\}}{{\rm Tr}\{\mathcal{M} I\}}$ for a  given
nucleon and electron helicity configuration.  An elementary calculation yields
the results quoted in~\S2.

As an exercise to illustrate the usefulness of these methods (and to prepare 
for the appearance of data on vector meson production in the final state of 
deep inelastic electron and muon scattering) I calculate the helicity 
information that can be extracted in production of vector mesons by polarized 
leptons scattering from unpolarized nucleons.

Since the decays of vector mesons such as the $\rho$, $K^*$ and $J/\psi$ 
conserve parity, it is impossible to measure their polarization from the decay 
angular distribution alone.  However, 
the decay angular distribution of a spin-one particle depends on the absolute 
value of 
its helicity.  Consider, for example, the $\rho$-meson decaying into 
$\pi\pi$.  In the $\rho$ rest frame (with the helicity defined along the $\hat 
e_3$-axis) the helicity zero state decays with a $\cos^2\theta$ distribution 
while the helicity $\pm 1$ states decay with a $\sin^2\theta$ distribution.

When classifying the independent fragmentation helicity amplitudes of a vector 
meson \cite{Ji}, one discovers several more beyond those present for 
spin-$\frac{1}{2}$.  The fragmentation (double) density matrix can be expressed 
in terms of the direct product of a $2\times 2$ quark density matrix and 
a $3\times 3$ meson density matrix with components along $(+,0,-)$ helicity.  
The result mimics eq.(\ref{cald}) with two exceptions.  (1) 
Time reversal allow two 
distinct helicity flip fragmentation functions \cite{Ji}; and (2) there is a 
``quadrupole'' fragmentation function analagous to the quadrupole distribution 
function called $b_1(x,Q^2)$ in Ref.~\cite{JM} which measures 
the difference in the quark fragmentation 
into the helicity zero and $\pm 1$ states of the meson.  It 
is this quadrupole fragmentation function that can be measured by scattering 
 electrons from an unpolarized nucleon target and observing the 
angular distribution of the vector meson decay products.  Ignoring 
Ji's possible T-violating fragmentation function (which requires nontrivial 
final state interactions and does not contribute to scattering from 
unpolarized targets), the fragmentation double density matrix for vector meson 
production is given by
\begin{equation}
{d{\mathcal {D}}\over dz} = \frac13 \hat 
q~I\otimes I  - \frac14 \hat b_q~I\otimes Q + \frac12 \Delta \hat q~
\vec\sigma\cdot\hat e'_3 \otimes
\vec S\cdot\hat e'_3 + \frac12 \delta \hat q~
\sum_{j=1,2}\vec\sigma\cdot\hat e'_j \otimes \vec S\cdot\hat e'_j 
\label{caldvec}
\end{equation}
where $\hat b_q$ is the quadrupole fragmentation function for a quark
of flavor $q$ to fragment to the vector meson, and $Q$ is the diagonal
member of the quadrupole helicity basis for spin-one, $Q={\rm
diag}(1,-2,1)$.  $\vec S$ are the spin-matrices for a spin-one
particle: $S_3={\rm diag}(1,0,-1)$, {\it etc.\/} As in the simple case
of spin-$1/2$, the structure of ${\mathcal{D}}$ can be read off the
definitions of the quark fragmentation functions in a helicity basis.

Combining this fragmentation density matrix with the distribution and hard 
scattering density matrices we find that $\hat b_q$ can contribute to 
scattering of unpolarized electrons from an unpolarized target as well as 
scattering of polarized electrons from a polarized target.  The relevant 
observable is the number of helicity-$0$ $\rho$'s minus the number of 
helicity-$\pm 1$ $\rho$'s divided by the sum.  For the case of unpolarized 
leptons and unpolarized target,
\begin{equation}
A_T\equiv {{N_0 - N_+ - N_-}\over 3(N_0 + N_+ + N_-)} = 
{\sum_q e_q^2 q_N(x,Q^2) \hat b_{q\rho}(z,Q^2)\over
\sum_q e_q^2 q_N(x,Q^2) \hat q_\rho(z,Q^2)}.
\end{equation}
A similar expression describes the production of $\rho$'s by a polarized beam from 
a polarized target.

\section{Acknowledgements}
I would like to thank Richard Milner, Michael D\" uren, Naomi Makins and 
members of the HERMES Collaboration for conversations.  
I am also grateful to Marek 
Karliner for comments, references and suggestions, and to the organizers of 
DIS'96 where this work was begun.

\end{document}